\documentclass[prd,twocolumn,eqsecnum,superscriptaddress,showpacs,letterpaper,altaffilletter]{revtex4}
\usepackage{color}
\usepackage{times}
\usepackage{graphicx}
\usepackage{fancyhdr}
\usepackage{float}

\renewcommand{\today}{\number\day\space\ifcase\month\or
  January\or February\or March\or April\or May\or June\or
  July\or August\or September\or October\or November\or December\fi
  \space\number\year}

\begin{document}

\title{Final Search for Lightly Ionizing Particles with the MACRO detector}

\begin{abstract}  
\vspace*{0.2in}

We present the final results of a search for lightly ionizing
particles using the entire cosmic ray data set the MACRO detector
collected during its 1995-2000 run.
Like the original search performed with the data of 1995-96,
this search was sensitive to fractionally charged particles
with an {\it{electric}} charge $q$ as low as  $\frac{e}{5}$
and with velocities between approximately $0.25c$ and $c$.
The efficiency of this search was $\approx 70 \,\%$ for 
$q = \frac{e}{5}$ and increased rapidly to $100 \,\%$ for higher charges.
No candidate events were observed.
This corresponds to a $90 \, \%$ C. L. upper limit on their isotropic
flux of
$6.1 \times 10^{-16} \,{\rm cm}^{-2}\,{\rm sec}^{-1} \,{\rm sr}^{-1}$
which represents the most stringent experimental limit ever obtained.
\end{abstract}


%\date[\relax]{ RCS \thercsid; compiled \today }
\date[\relax]{ compiled \today }

\author{
{\rm The MACRO Collaboration}\\
M.~Ambrosio$^{12}$, 
R.~Antolini$^{7}$, 
A.~Baldini$^{13}$, 
G.~C.~Barbarino$^{12}$, 
B.~C.~Barish$^{4}$, 
G.~Battistoni$^{6,a}$, 
Y.~Becherini$^{2}$,
R.~Bellotti$^{1}$, 
C.~Bemporad$^{13}$, 
P.~Bernardini$^{10}$
% ...
\\
\vspace{1.5 cm}
\footnotesize
1. Dipartimento di Fisica dell'Universit\`a  di Bari and INFN, 70126 Bari, 
 Italy \\
2. Dipartimento di Fisica dell'Universit\`a  di Bologna and INFN, 40126 
Bologna, Italy \\
3. Physics Department, Boston University, Boston, MA 02215, USA \\
4. California Institute of Technology, Pasadena, CA 91125, USA \\
5. Department of Physics, Drexel University, Philadelphia, PA 19104, USA \\
6. Laboratori Nazionali di Frascati dell'INFN, 00044 Frascati (Roma), Italy \\
7. Laboratori Nazionali del Gran Sasso dell'INFN, 67010 Assergi (L'Aquila), 
 Italy \\
8. Depts. of Physics and of Astronomy, Indiana University, Bloomington, IN 
47405, USA \\
9. Dipartimento di Fisica dell'Universit\`a  dell'Aquila and INFN, 67100 
L'Aquila, Italy\\
10. Dipartimento di Fisica dell'Universit\`a  di Lecce and INFN, 73100 Lecce, 
 Italy \\
11. Department of Physics, University of Michigan, Ann Arbor, MI 48109, USA \\
12. Dipartimento di Fisica dell'Universit\`a  di Napoli and INFN, 80125 
Napoli, Italy \\
13. Dipartimento di Fisica dell'Universit\`a  di Pisa and INFN, 56010 Pisa, 
Italy \\
14. Dipartimento di Fisica dell'Universit\`a  di Roma "La Sapienza" and 
INFN, 00185 Roma, Italy \\
15. Physics Department, Texas A\&M University, College Station, TX 77843, 
 USA \\
16. Dipartimento di Fisica Sperimentale dell'Universit\`a  di Torino and 
INFN, 10125 Torino, Italy \\
17. L.P.T.P, Faculty of Sciences, University Mohamed I, B.P. 524 Oujda, 
 Morocco \\
18. Dipartimento di Fisica dell'Universit\`a  di Roma Tre and INFN Sezione 
Roma Tre, 00146 Roma, Italy \\
$a$ Also INFN Milano, 20133 Milano, Italy \\
$b$ Also IASF/CNR, Sezione di Bologna, 40129 Bologna, Italy \\
$c$ Also Department of Physics, Pennsylvania State University, University 
Park, PA 16801, USA \\
$d$ Also Universit\`a  di Trieste and INFN, 34100 Trieste, Italy \\
$e$ Also U. Minn. Duluth Physics Dept., Duluth, MN 55812 \\
$f$ Also Dipartimento di Energetica, Universit\`a  di Roma, 00185 Roma, 
 Italy \\
$g$ Also Dept. of Physics, MIT, Cambridge, MA 02139 \\
$h$ Also Intervideo Inc., Torrance CA 90505 USA \\
$i$ Also Dipartimento di Fisica dell'Universit\`a  della Calabria, Rende 
(Cosenza), Italy \\
$j$ Also RPD, PINSTECH, P.O. Nilore, Islamabad, Pakistan\\
$k$ Also Institute for Space Sciences, 76900 Bucharest, Romania \\
$l$ Also Universit\`a  della Basilicata, 85100 Potenza, Italy \\
$m$ Also Dipartimento di Scienze e Tecnologie Avanzate, Universit\`a  del 
Piemonte Orientale, Alessandria, Italy \\
$n$ Also Resonance Photonics, Markham, Ontario, Canada\\}
%
%\author{Author List}
%\affiliation{MACRO Collaboration}
%\input{authorList}

\maketitle

%%%%%%%%%%%%%%%%%%%%%%%%%%%%%%%%%%%%%%%%%%%%%%%%%%%%%%%%%%%%%%%%%%

\section{Introduction}
\label{sec:intro}

Within the Standard Model of electroweak and strong interactions
electric charge quantization remains unexplained.
Since Millikan's oil drop experiment demonstrated
electric charges in Nature come in discrete units,
numerous experimenters have refined the original determination of
the unit charge ({\it{e}}) and searched for free particles with
fractional charge.
This interest was intensified with the proposal
of the  quark model by Gell-Mann~\cite{ge64} and Zweig~\cite{zw64}
in $1964$ and its undisputed success in explaining the deep
inelastic scattering experiments in the late 1960's~\cite{slac}.
This model --now part of the Standard Model-- does
require that single quarks come with a fractional charge, however,
it does not allow them to exist as free particles.
Quarks are confined in color-neutral baryons and mesons that
carry integer electric charges.

Searches for fractionally charged particles 
have thus been carried out over the last decades
in bulk matter, at accelerators and in cosmic rays
~\cite{laura,sm89,kl95,ly85,jo77,ha00}
without any evidence for their existence.
The observation of such particles
would be direct evidence of physics beyond the Standard Model.
Fractionally charged particles are easily accomodated  
in Grand Unified Theories (GUT) of the electroweak and strong 
interactions.
Within this framework the quantization of the electric charge 
is also explained as a consequence of the 
non trivial commutation relations between the operators in the theory.
Simple extensions of the {\bf SU(5)} unification group allow for 
$\frac{e}{3}$ and $\frac{2e}{3}$ charges~\cite{fr82,ba83} while
larger groups such as {\bf SU(8)}~\cite{yu84}, {\bf SO(14)}~\cite{ya83} 
and {\bf SO(18)}~\cite{do83} allow for 
$\frac{e}{2}$ ones.
Many popular superstring models~\cite{we85} 
yield stable particles with charges down to $\frac{e}{5}$ or even smaller. 
Finally, some spontaneously broken 
QCD theories predict the possibility of primordial unconfined 
quarks and gluons~\cite{de78} contained in superheavy quark-nucleon 
complexes with large noninteger charge. 
 
Previous searches for fractionally charged particles in the 
penetrating cosmic radiation include the search using the first
year of MACRO's running ~\cite{ma00lip} as
well as searches
with the Kamiokande-II~\cite{ka91} and LSD~\cite{ls94} experiments.
Until this search the best $90 \,\%$ C. L. upper flux limits
were obtained by the Kamiokande experiment and
are $2.1$ and $2.3 \times 10^{-15} \,{\rm cm}^{-2} \,{\rm sec}^{-1} 
\,{\rm sr}^{-1}$ for charges $\frac{e}{3}$ and $\frac{2e}{3}$, respectively.
Here we present the final results of the search for particles with
fractional charge from $\frac{e}{5}$ to $\frac{2e}{3}$
in the penetrating cosmic 
radiation using the entire dataset of the MACRO experiment. 
The search method is based on the assumption of a reduced
ionization and atomic excitation energy loss rate of a
fractionally 
charged particle with respect to that of a unit charge.
Since the energy 
loss rate of a particle of charge $Q$ is proportional to $Q^2$, a 
fractionally charged particle is expected to lose energy, by 
excitation and ionization, at a much lower rate than a minimum 
ionizing particle of the same velocity.
The fractionally charged 
particles are therefore also called {\it Lightly Ionizing Particles}
(LIPs).
This search would not have been sensitive to fractionally charged
particles that may interact also strongly (e.g., standard model
quarks) as these particles would not be able to penetrate large
amounts of material.

%%%%%%%%%%%%%%%%%%%%%%%%%%%%%%%%%%%%%%%%%%%%%%%%%%%%%%%%%%%%%%%%%%

\section{Experimental Setup}
\label{sec:setup}

The MACRO experiment, described in detail in~\cite{ma93tec,ma02tec}, was 
a multipurpose underground detector located in Hall B of 
the Gran Sasso National Laboratory of the Italian Institute of
Nuclear Physics (INFN).
It was optimized for the search 
for magnetic monopoles~\cite{ma02mon,ma02cat}
but it also presented excellent capabilities for
studying atmospheric neutrino oscillations~\cite{ma95nos,ma98nos,ma01nos},
cosmic rays~\cite{ma99mu,ma02mu}, astrophysical 
point sources~\cite{ma01nast,ma02nast}, neutrinos from 
gravitational collapses~\cite{ma92col,ma98col}.
The detector had a modular structure; it was  
composed of six supermodules each of which was
divided into a lower and an upper part (\lq\lq attico\rq\rq).
All supermodules were equipped with three detector sub-sys\-tems: limited 
streamer tubes, liquid scintillator counters
and nuclear track detectors.
The overall dimensions of the  apparatus were
$76.6 \times 12 \times 9.3~{\rm m}^3$ and the acceptance for 
an isotropic flux of particles was $\approx 10^4~{\rm m}^2 \, {\rm sr}$. 
The detector was active in various different configurations
from the fall of $1989$ 
until December $2000$.

\begin{figure}[!thb]
\includegraphics[width=0.95\linewidth]{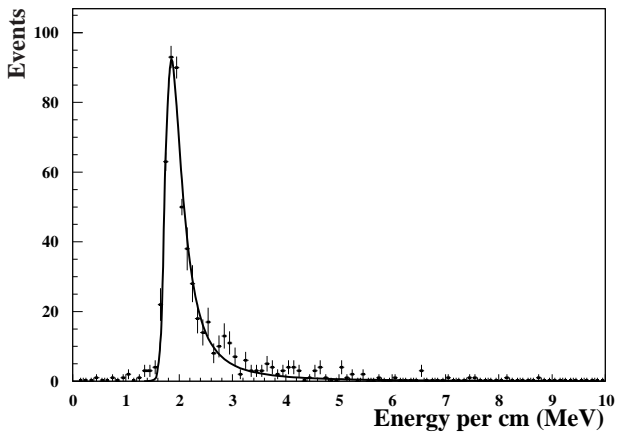}
\caption{
Typical energy loss rate in a MACRO liquid scintillator 
counter.
The superimposed fit is a Landau distribution folded with 
the photoelectron fluctuations.
One may notice the low $dE/dx$ tail of the energy loss
distribution that extends within the LIP signal region.
}
\label{fig:landau}
\end{figure}

\begin{figure}[!thb]
\includegraphics[width=0.95\linewidth]{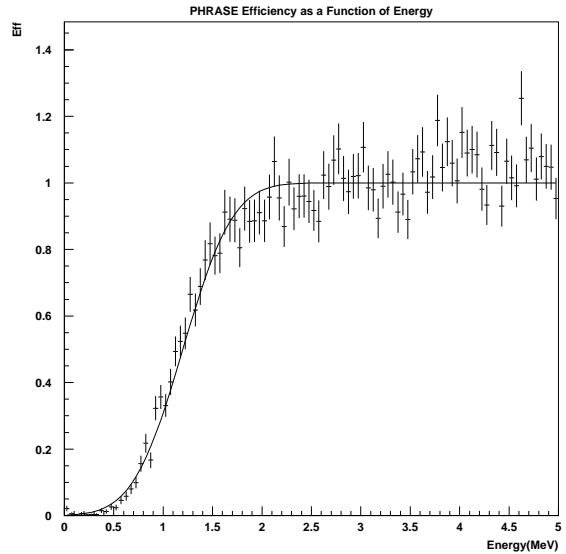}
\caption{
Measured trigger efficiency of the low-energy PHRASE 
trigger (i.e., LIP trigger) as a function of the energy released in the 
liquid scintillator counters. Some measured efficiencies were larger 
than $100 \, \%$ because the normalization factor used was an estimate of 
the true normalization as a function of energy.
}
\label{fig:phreff}   
\end{figure}

Lightly ionizing tracks in MACRO would be signatures for fractionally
charged particles.
They could be identified both in the streamer tube and the scintillator system.
The streamer tube system provided a three-dimensional reconstruction of
the particle's trajectory.
This system was more than $99\%$ efficient in forming a track upon the passage
of a fractional charge $\frac{1}{5}e$ or greater~\cite{ma00lip,walter}.
The scintillator system provided redundant measurements of the particle's
energy loss.
Time-of-flight information with sub-nanosecond accuracy was also
recorded.
A Lightly Ionizing Particle (LIP) trigger was formed in hardware
by combining
MACRO's primary streamer tube trigger ~\cite{ma02tec,walter,ma00lip}
with a coincidence of MACRO's lowest energy threshold
scintillator trigger, called PHRASE ~\cite{ma93tec,ma02tec,ma92col}, 
coming from at least three detector faces.
The coincidence window of $400~{\rm ns}$ among the scintillator face
triggers was setting the minimum particle velocity the trigger
hardware was sensitive to.
This was at about $0.25c$ for most of the detector's acceptance.
PHRASE provided a trigger based on a single
scintillator counter
with a threshold at approximately $1.2~{\rm MeV}$.
Given that  the  Landau peak for cosmic ray muons 
was approximately $38~{\rm MeV}$ for a typical $20~{\rm cm}$ pathlength
within a scintillator counter (see fig. \ref{fig:landau})
the $1.2~{\rm MeV}$ energy
threshold set a minimum fractional charge of $\frac{1}{5}e$
that MACRO could detect.
A typical plot of
the measured trigger efficiency of the low-energy PHRASE 
trigger as a function of energy is shown in fig. \ref{fig:phreff}
as obtained with the 1995-96 data \cite{ma00lip}.
The low energy threshold of the PHRASE was periodically
checked throughout the five-year run
using natural radioactivity decay lines and it was maintained
consistently at the $1-1.2~{\rm MeV}$ level during
the entire run.

The LIP circuitry provided the hit pattern information of the scintillator
counters that exceeded the $1.2~{\rm MeV}$ threshold.
For all counters involved in a LIP trigger their photomultiplier
(PMT) wave forms recorded on MACRO's $200~{\rm MHz}$ wave form digitizer
(WFD) system were read.
An analysis of these wave forms could then provide the localization
of the event along the scintillator counter as well as the reconstruction
of the energy released.
Both of these when combined with the
three-dimensional track information provided by the streamer tubes
could effectively reject any radioactivity or cosmic ray muon signal
at a level of better than $1/10^{6}$ leaving a very low background
search for LIPs.
This was the approach in our first search for LIPS
using MACRO's 1995-96 data~\cite{ma00lip}.

%%%%%%%%%%%%%%%%%%%%%%%%%%%%%%%%%%%%%%%%%%%%%%%%%%%%%%%%%%%%%%%%%%

\section{Data Analysis}
\label{sec:macrolip}
 
The MACRO scintillator subdetector was equipped with three independent systems
able to record the time of flight, position and energy
loss of a lightly ionizing particle as it crossed the apparatus:
the primary muon trigger (called ERP), the primary gravitational
collapse trigger (called PHRASE) and the Wave-Form-Digitizer
system (WFD).
While all three had comparable resolutions
~\cite{ma02tec,ma98nos,ma98col,ma93tec,ma02tec,ma00lip}
for minimum ionizing
particles 
their different thresholds implied sensitivity to as low as
$\sim\frac{3e}{5}$ for the ERP (given its $\sim15~{\rm MeV}$ threshold),
$\sim\frac{2e}{5}$ for the PHRASE (given its $\sim7~{\rm MeV}$ threshold),
and $\sim\frac{e}{5}$ for the LIP-triggered WFD system.
The LIP trigger was designed to trigger on cosmic ray muons in addition to lightly ionizing particles.
We have thus used the ERP and PHRASE systems in order to identify and
reject the cosmic ray muon component that was present in our
LIP trigger data set.
This allowed the wave form digitizer
system to be invoked
only when the ERP and PHRASE had failed to record a candidate
event presumably because of their energy thresholds and triggering inefficiencies relative to LIP/WFD system.
The energy thresholds of the PHRASE and ERP systems were
continuously monitored using natural radioactivity and cosmic ray muons. They  were periodically adjusted in order to account for any electronic drifts or photomultiplier tube gains.
This guaranteed their consistency and maintained
 our ability to trigger and reconstruct
tracks down to $\frac{e}{5}$ charges throughtout the entire data
taking period.

The data set for this search comes from the 5-year run of the
MACRO detector in its final configuration from July 1995
to December 2000. 
Run quality criteria were first applied; they required the
MACRO detector to be running in its full configuration (i.e.,
all six supermodules, excluding counters with calibration
errors) and without any serious acquisition
problems (e.g., high dead time, repeated hardware errors,
high voltage power supply glitches or high voltage 
errors what-so-ever).
The integrated live-time was $1320$ days and during that period
$18.3\times10^6$
LIP triggers were collected.
The analysis of these triggers proceeded as follows:
\begin{itemize}
\item[1)] 
      Using the streamer tube hit information, we first required a single
      streamer tube track reconstructed.
      In addition, 
      using the LIP scintillator counter hit information,
      we required hits to be present in no 
      more than four scintillator faces and within six scintillator
      counters in the same face.
      Given the geometry of the detector, no single candidate LIP track
      could result in such a scintillator counter hit pattern.
      This cut primarily rejected cosmic ray muons 
      accompanied by electromagnetic showers.
\item[2)]
      Using the streamer tube track parameters
      we reconstructed the event longitudinal
      position along the scintillator counters
      as well as the path length of the crossing particles in them.
      We required the path length to be between 
      $13~{\rm cm}$ and $70~{\rm cm}$ and the hit position along
      the counter within the central $10.8~{\rm m}$ part of it
      (i.e., rejecting tracks within the final $10~{\rm cm}$ of a scintillator
      counter).
      Both these cuts eliminated track geometries that were more susceptible
      to energy reconstruction errors either due to low photoelectron
      statistics (short pathlengths) or to poorly calibrated
      PMT responses (near-PMT geometric response).
      When available, we required that the position along the counter
      provided by the scintillator timing was in agreement within
      $80~{\rm cm}$ ($6 \,\sigma$ )
      with the  position reconstructed by the streamer
      tube system.
      Following these geometrical cuts, the detector acceptance was about 
      $3300~{\rm m^2 \, sr}$ for an isotropic flux of particles.
\item[3)]
      For every LIP counter that was intercepted by a track that fulfilled
      all cuts so far, we assigned the energy loss as calculated for
      the same scintillator counter(s)
      by the ERP muon system in the absence of which
      this information was sought in the PHRASE gravitational collapse
      system.
      In lack of any energy loss information from these two systems
      a {\it{zero}} energy loss was assigned to that hit;
      this makes
      {\it{a priori}} the best LIP candidate event in MACRO.
\end{itemize}

\begin{figure}[!thb]
\includegraphics[width=0.95\linewidth]{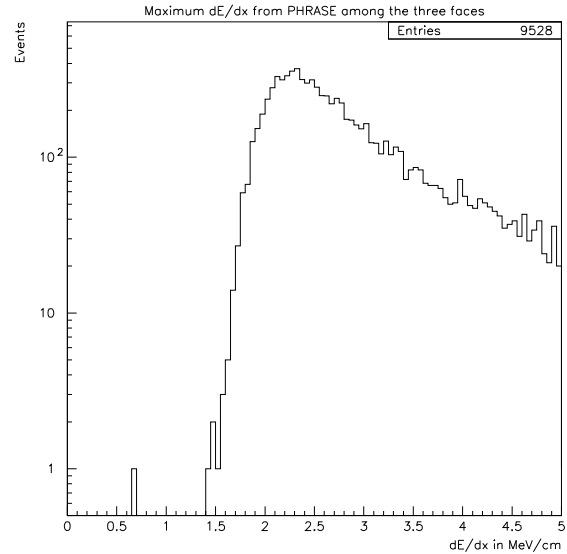}
\caption{
Energy loss as measured by PHRASE for the LIP 
events that passed the track quality and geometry cuts and satisfied 
the requirement of a maximum energy loss rate (measured by ERP) 
less than $1.1~{\rm MeV/cm}$.
This dataset was comprised primarily of events where the ERP system
did not trigger at all due to inefficiencies.
For these events an energy measurement was sought in the PHRASE system.
The signal region is in the 
$\left[0, 1.35\right]~{\rm MeV/cm}$ interval. 
}
\label{fig:anaeloss}   
\end{figure}

\begin{figure}[!thb]
\includegraphics[width=0.95\linewidth]{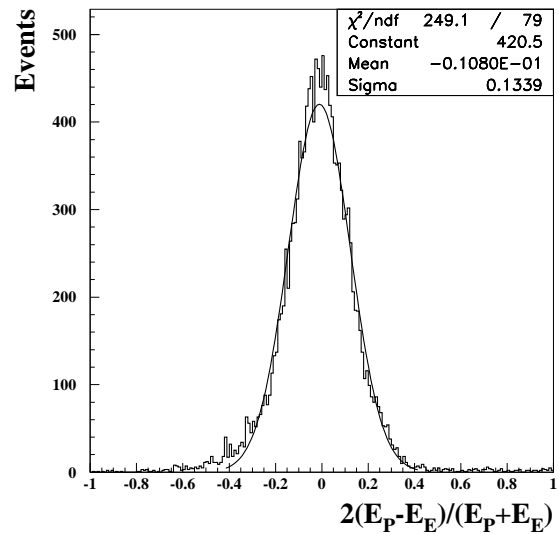}
\caption{
Comparison between the PHRASE and ERP measured 
energies in the range of $10 \div 100~{\rm MeV}$. 
The distribution of the difference between the two 
energies normalized to their average shows how
well these two agree.
The distribution is roughly gaussian with $\sigma = 0.13$.
}
\label{fig:phrerpcomp}   
\end{figure}

This last stage of the LIP analysis enabled us to perform
a measurement of the particle energy loss rate $dE/dx$
for all cases that the LIP event was accompanied by a 
muon and/or the gravitational collapse trigger.
The maximum $dE/dx$  among the counters involved is plotted
in figure \ref{fig:anaeloss}.
All tracks in this analysis were
accompanied by a muon and/or gravitational collapse hit.
This allowed us to establish the particle's
$dE/dx$ without making use of the wave form digitizer system.
Since energy information from both the PHRASE and ERP systems
was used in this search, we verified that in the relevant
energy range ($10 \div 100~{\rm MeV}$) the
two systems agreed to within $20\,\%$ or better in more
than  $95 \, \%$ of the liquid scintillator counters
(see fig.~\ref{fig:phrerpcomp}).

%%%%%%%%%%%%%%%%%%%%%%%%%%%%%%%%%%%%%%%%%%%%%%%%%%%%%%%%%%%%%%%%%%

\section{Results and Discussion}
\label{sec:resdisc}

The {\it{expected signal region}} for this LIP search was set below
$1.1~{\rm MeV/cm}$ for the maximum energy loss rate --when measured
by the ERP-- in any of
the scintillator counters intercepted by a streamer tube track.
Since the energy loss was measured by PHRASE whenever ERP
information was not available
and in order to account for
the potential $20\%$ mismatch in their energy estimates,
the signal region was extended up to $1.35~{\rm MeV/cm}$ 
whenever a measurement was performed by PHRASE.
Above this level the cosmic ray muon energy loss
spectrum does not permit any identification of fractionally
charged particles thus setting an upper limit of sensitivity to
approximately $\frac{2e}{3}$.

As one can see in fig.~\ref{fig:anaeloss}
there is one event (run 15871, event 5649)
that appears in the signal region.
It corresponds 
to a maximum energy loss of 
$0.66~{\rm MeV/cm}$, i.e., about $20 \,\%$ lower 
than what expected for a particle of charge $\frac{2e}{3}$ and about a factor 
of $3$ higher than what expected for a particle of charge $\frac{e}{3}$. 
Three 
scintillator counters were involved in this trigger; the first 
in one of the upper vertical layers, the second in the central horizontal 
layer and the third in the lower horizontal layer.
There were no ERP 
triggers involved (suggesting the energy released in each of 
the counters was below $15~{\rm MeV}$) and only one PHRASE hit in the vertical
layer was present (suggesting the energy released in the other two counters
was below $7~{\rm MeV}$).
The energy loss measured by PHRASE for this hit was $13.7~{\rm MeV}$ and 
using the path length in the box provided by the tracking
we computed a maximum energy loss rate of $0.66~{\rm MeV/cm}$. 
The position along the counter for this particular box measured by the
PHRASE and by the streamer tube track geometry
were in agreement (within $15~{\rm cm}$).
We have examined this event by hand relying primarily on the
wave forms as recorded for all the counters involved in the
trigger.
The apparent amplitude of the recorded wave forms was consistent
with the energy thresholds for the ERP and the PHRASE.
Having three scintillator counters involved in the trigger
we have checked for a consistency in the relative
timing of them with the crossing of a single particle
of constant velocity.
The relative timing between the counter
in the upper part of the detector and that in the central part was 
consistent with the passage of a relativistic particle coming
from above while
the relative timing between the box in the lower part of the
detector and any of the other two hits was 
consistent with a slowly moving upward-going particle.
We thus discarded this event from the signal region.

%%%%%%%%%%%%%%%%%%%%%%%%%%%%%%%%%%%%%%%%%%%%%%%%%%%%%%%%%%%%%%%%%%

\section{Conclusions}
\label{sec:flux}

Over the five years of running of the MACRO detector
in its final configuration (1995-2000)
we observed no candidates for LIPs in cosmic ray single tracks.

Given the detector's acceptance and live time
we computed an integrated exposure of
$3.8 \times 10^{15} \, {\rm cm^2 \, s \, sr}$
which for an isotropic flux of particles yields a $90\%$
C.L. upper flux limit of
$6.1 \times 10^{-16} \, {\rm cm^{-2} \, s^{-1} \, sr^{-1}}$.
The search was fully efficient
for charges $\frac{e}{4}$ to $\frac{2e}{3}$.
The flux limit rises to higher values for lower charges
as the detection efficiency drops from $100\%$ to lower values.
This limit improves the previously published MACRO
result~\cite{ma00lip} by over an order of magnitude.
The previously published MACRO search applies to LIPs
present in both single and multi-track events while this
one (as well as the ones from experiments other than MACRO)
applies to LIPs present in single tracks only.

\begin{figure}[!thb]
\includegraphics[width=0.95\linewidth]{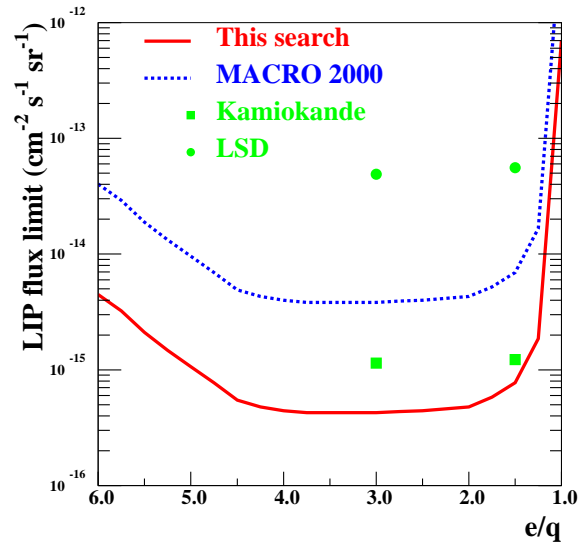}
\caption{
The $90 \, \%$ C.L. MACRO flux upper limits for LIPs 
(solid line) compared with previous limits from MACRO, LSD 
and Kamiokande.
}
\label{fig:limits}
\end{figure}

This MACRO result
can be compared with those obtained by the Kamiokande~\cite{ka91} 
and the LSD~\cite{ls94} experiments keeping in mind
that all these are limits on the local flux of LIPs at the detector site.
Kamiokande and LSD quoted their 
results for two specific values of fractional charge, namely,
$\frac{e}{3}$ and $\frac{2e}{3}$, while the MACRO result applies to a
continuum of charges.
It should also be noted that
all limits were obtained with the assumption
of an isotropic flux of LIPs.
This is reasonable if we assume that 
LIPs are produced in the rock around the detector (for instance,
by cosmic ray muon interactions or some other unknown mechanism).
However, if LIPs are produced by the cosmic ray interactions
in the upper atmosphere or impinge on the earth from outer space, the 
assumption of an isotropic flux is no longer valid.
In this case a detailed physical model for the production and
propagation of LIPs as well as of the response of the detector
would be needed in order to derive and compare results for
the various experiments.
In the most obvious case, LIPs without enough initial kinetic
energy will not be able to reach the detector from directions
below the horizon thus worsening all upper limits by a factor
of two.

%%%%%%%%%%%%%%%%%%%%%%%%%%%%%%%%%%%%%%%%%%%%%%%%%%%%%%%%%%%%%%%%%%

\begin{acknowledgments}
We gratefully acknowledge the support of the director and of the staff 
of the Laboratori Nazionali del Gran Sasso and the invaluable assistance 
of the technical staff of the Institutions participating in the experiment. 
We thank the Istituto Nazionale di Fisica Nucleare (INFN), the U.S. 
Department of Energy and the U.S. National Science Foundation for their 
generous support of the MACRO experiment. We thank INFN, ICTP (Trieste) 
and World Laboratory for providing fellowships and grants for non Italian 
citizens.   
\end{acknowledgments}

\end{document}